\documentclass[aps,pra,reprint,onecolumn]{revtex4-1}
\pdfoutput=1
\usepackage[utf8]{inputenc}
\usepackage[T1]{fontenc}
\usepackage{graphicx}
\usepackage{amsmath,amsfonts,amssymb}
\usepackage{bm}             % \bm
\usepackage{placeins}       % \FloatBarrier

\usepackage{hyperref}
\hypersetup{
    bookmarks=true,
    pdftitle={Many-body excitations and de-excitations in trapped ultracold bosonic clouds},
    pdfauthor={Marcus Theisen, Alexej Streltsov},
    pdfkeywords={MCTDHB,BEC,many-body physics,dynamic control},
    colorlinks=false,
}

\def\r  {\bm{\mathrm{r}}}
\def\R  {\bm{\mathrm{R}}}

\def\u  {\bm{\mathrm{u}}}
\def\v  {\bm{\mathrm{v}}}
\def\C  {\bm{\mathrm{C}}}

\def\bcalL {\bm{\mathcal{L}}}
\def\brho  {\bm{\rho}}
\def\bphi  {\bm{\phi}}

\DeclareMathOperator\Var  {Var}
\DeclareMathOperator\E    {E}

\def\fm {\footnotemark}

\begin{document}
\title{Many-body excitations and de-excitations in trapped ultracold bosonic clouds}

\author{Marcus Theisen}
\email{m.theisen@stud.uni-heidelberg.de}
\author{Alexej I. Streltsov}
\email{alexej.streltsov@pci.uni-heidelberg.de}
\affiliation{
  Theoretische Chemie,
  Physikalisch-Chemisches Institut,
  Universität Heidelberg,
  Im Neuenheimer Feld 229,
  D-69120 Heidelberg,
  Germany
}
\date{\today}

\begin{abstract}
We employ the {M}ulti{C}onfiguraional {T}ime-{D}ependent {H}artree for {B}osons (MCTDHB) method to study excited states of interacting Bose-Einstein condensates confined by harmonic and double-well trap potentials.
Two approaches to access excitations, a static and a dynamic one, have been studied and contrasted.
In static simulations the low-lying excitations have been computed by utilizing the LR-MCTDHB method - a linear response theory constructed on-top of a static MCTDHB solution.
Complimentary, we propose two dynamic protocols that address excitations by propagating the MCTDHB wave-function.
In particular, we investigate dipole-like oscillations induced by shifting the origin of the confining potential and breathing-like excitations by quenching frequency of a parabolic part of the trap.
To contrast static predictions and dynamic results we have computed time-evolutions and their Fourier transforms of several local and non-local observables.
Namely, we study evolution of the $\langle x(t) \rangle$, its variance $\operatorname{Var}(x(t))$, and of a local density computed at a selected position.
We found out that the variance is the most sensitive and informative quantity - along with excitations it contains information about the de-excitations even in a linear regime of the induced dynamics.
The dynamic protocols are found to access the many-body excitations predicted by the static LR-MCTDHB approach.
\end{abstract}

\maketitle

\section{\label{sec:intro}Introduction}
Many-body approaches are crucial for a modern understanding and physical description of ultracold quantum gases.
Particularly, systems of strongly interacting ultracold atoms demand for theories that predict dynamic behavior and excitation spectra~\cite{All_review0,All_review1,All_review2, Pethick}.
A system of major interest is the Bose-Einstein condensate (BEC). With its experimental realization~\cite{anderson95,davis95} a well-controllable mesoscopic quantum object is at hand offering ways to study many-body systems and excitations therein empirically.
Experimentally, owing to their controllability, excitations of BECs have been widely explored~\cite{folman02,Proukakis,Ozeri,Shammass}.
At the same time tremendous theoretical efforts seek to account for the observed correlation phenomena and their proper many-body descriptions~\cite{All_review0,All_review1,All_review2, Pethick}.
Since analytic solutions to many-body problems are rare, numerical simulations are inevitable.

The most popular mean-field theory to treat BECs in dilute, ultracold vapors is based upon the Gross-Pitaevskii (GP) equation ~\cite{Pethick}.
Within the GP approach all constituting bosons are assumed to occupy a single one-particle state, describing thereby  a simple, i.e., fully condensed and coherent condensate \cite{Penrose}.
Clearly, whenever the BEC is partially condensed, depleted, or multi-fold fragmented \cite{StJames} a single wave-function approach is insufficient.
On physical side all these phenomena originate from strong interparticle interactions and multi-well topologies of the confining traps.
Experimentally, connections between strong interactions and quantum depletion have been established in ~\cite{xu06}, the fragmentation phenomena in double-well traps have been observed in \cite{Kasevich0,JoergNatPhys06}.
First experimental realization of the Mott insulator states in multi-well traps formed by optical lattices has been reported in ~\cite{greiner02}.
On theoretical side to grasp depletion and fragmentation phenomena one has to go beyond the one-mode GP ansatz and use more complicated many-body theories where the bosonic wave-function is constructed with multiple (one-particle) orbital modes \cite{menotti01,frag2,frag3,frag4,frag5}.
In this paper to describe condensation, depletion and fragmentation phenomena in static setups and time-dependent processes on the same ground we use the Multi-Configurational Time-Dependent Hartree method for Bosons (MCTDHB) ~\cite{Streltsov07,Alon08}, which is available within the MCTDHB-Laboratory package~\cite{MCTDHB-Lab}. 

The standard static, i.e., time-independent approach to calculate excitation spectra of BECs at the one-orbital mean-field level is to apply the Linear Response (LR) theory to the GP ground state.
The underlying equations are often called Bogoliubov-de Gennes (BdG$\;\equiv\;$LR-GP) equations~\cite{bogolubov47a}.
Recently, the same linear-response idea was applied to the multi-orbital best mean-field ~\cite{grond12} and many-body MCTDHB ~\cite{grond13} ground states allowing thereby to access excitations of the multi-fold fragmented BECs at the mean-field and many-body levels. 
The invented~\cite{grond13}, developed~\cite{MCTDHB-Lab}, generalized~\cite{LRuni} and benchmarked~\cite{LRgeneralbench} LR-MCTDHB theory has allowed us to discover low-lying excitations in trapped systems which are not described within the BdG theory.
As one might expect such states are present in fragmented systems~\cite{grond12} and, surprisingly, also in condensed systems~\cite{grond13,LRuni,LRgeneralbench} where BdG was believed to govern the physics.

Naturally, questions of interest concern the origin of many-body excited states, their properties and possible classification schemes.
Here, we attempt to generalize characterization of the excitations obtained in the non-interacting case to the interacting systems.
We also distinguish two kinds of many-body excitations - one branch describes collective excitations while the second one represents excitations involving multiple particles.
These excitations will be defined and described in some detail below.

Another issue tackled in this paper is a dynamic control.
The long-term perspective is to find dynamic protocols that allow for a controlled population of the desired excitations.
Within the framework of the MCTDHB method this problem can, in principle, be solved  by applying the optimal-control theory as proposed in ~\cite{OptControl1,OptControl2,OptControl3,OptControl4} and, more recently, by merging the general optimal-control Chopped RAndom Basis (CRAB) algorithms~\cite{CRAB1, CRAB2} with the MCTDHB method~\cite{CRAB-MCTDHB}.
Here, however, we present and consider two simple protocols that excite BECs via sudden modification of the system's Hamiltonian.
Such manipulations manifest in oscillation of the many-body wave-function, and thus the particle density, in time.
In an experimental setup changes on the Hamiltonian can be realized by introducing control fields, altering the confining potential or toggling the inter-particle interaction using Feshbach resonances~\cite{feshbach}.
An open question is how to detect and verify many-body excitations confidently.
A first step towards the answer is to find out which observables provide more detailed information about the many-body excitations.
To this end, we propose to use the positional variance as a sensitive probe for many-body excitations and de-excitations.

The structure of this paper is as follows.
In Sec.~\ref{sec:system} we introduce the systems of interest by giving the Hamiltonian and specifying the external trapping potential.
Sec.~\ref{sec:results} contains numerical results of BECs described at mean-field (GP) and many-body (MCTDHB) levels.
Two methods to analyze excitation spectra are discussed and compared:
A static approach using linear response theory (Sec.~\ref{sec:statics}) and a dynamic approach using wave-packet propagation (Sec.~\ref{sec:dynamics}).
Finally, in Sec.~\ref{sec:discussion} we summarize the results and give future prospects.

\section{\label{sec:system}System and Hamiltonian}
The MCTDHB($M$) algorithm~\cite{Streltsov07,Alon08} effectively solves the time-dependent many-body Schr\"odinger equation
\begin{equation}\label{eq:TDSE}
\hat{H}\Psi
= i\hbar \frac{\partial}{\partial t}\Psi \,,
\end{equation}
taking the sum over symmetrized Hartree products (or \textit{permanents}) as the ansatz for the many-body wave-function
\begin{equation}\label{eq:Psi}
\left|\Psi(t)\right>
= \sum_{\vec{n}}C_{\vec{n}}(t)\left|\vec{n};t\right> \,.
\end{equation}
Here, the summation runs over all possible configurations of the state vector $\vec{n} = (n_1,n_2,\ldots,n_M)$ that preserve the total number of bosons $N=\sum_{i}^{M}n_{i}$ and $M$ denotes the number of one-particle wave-functions $\phi_{i}(\r,t)$ (or \textit{orbitals}) used to construct the respective permanents.

The shape of the orbitals $\phi_{i}(\r,t)$ and the expansion coefficients $C_{\vec{n}}(t)$, which account for normalization, are variational time-dependent parameters of the MCTDHB method.
We would like to stress that the single orbital MCTDHB($M=1$) approach is fully equivalent to the Gross-Pitaevskii theory.
Using the MCTDHB($M$) method with $M\ge2$ orbitals implies an above mean-field, many-body treatment.

Given the MCTDHB wave-function at every time-slice $t$ we can construct and diagonalize the reduced one-body density matrix $\rho(\r_1,\r_1';t)$.
It can be expressed in terms of the \textit{natural orbitals} and occupations, i.e., its eigenstates $\phi^{NO}_i(\r,t)$ and eigenvalues $n_i(t)$:
\begin{align}\label{eq:dns}
\rho(\r_1,\r_1';t)
&\equiv
N \int \Psi^*(\r_1',\r_2,\dots,\r_N,t)\: \Psi(\r_1,\r_2,\dots,\r_N,t)\:
  \mathrm{d}\r_2\cdots\mathrm{d}\r_N \nonumber\\
&=
\sum^M_{i,j} \rho_{ij}(t)\: \phi^*_j(\r_1',t)\: \phi_i(\r_1,t) \nonumber\\
&=
\sum^M_{i} n_{i}(t)\: \phi^{\ast NO}_i(\r_1',t)\: \phi^{NO}_i(\r_1,t) \,.
\end{align}

The many-body state is called condensed if only one natural eigenfunction -- \textit{condensate orbital} is $\approx100\%$ occupied.
The BEC may be called depleted when the non-condensed fraction, i.e., the total occupation of other than the condensate orbital ($\sum^M_{i>1}n_i$) is of the order of $10\%$.
When several eigenfunctions have macroscopic occupation the system is called fragmented.
For completeness we also present the expression for the two-body density, which we will use later:
\begin{align}\label{eq:dns2}
\rho(\r_1,\r_2,\r_1',\r_2';t)
&\equiv
N(N-1) \int \Psi^*(\r_1',\r_2',\dots,\r_N,t)\: \Psi(\r_1,\r_2,\dots,\r_N,t)\:
  \mathrm{d}\r_3\cdots\mathrm{d}\r_N \nonumber\\
&=
\sum^M_{i,j,k,l} \rho_{ijkl}(t)\:
  \phi^*_j(\r_1',t)\: \phi^*_l(\r_2',t)\: \phi_i(\r_1,t)\: \phi_k(\r_2,t) \,.
\end{align}

For $N$ interacting particles the Hamiltonian reads
\begin{equation}\label{eq:ham}
\hat{H}
= \sum_{j=1}^N \hat{h}(\r_j)
+ \sum_{j<k}^N \lambda_0\: W(\r_j - \r_k) \,,
\end{equation}
with the one-particle Hamiltonian
\begin{equation}\label{eq:ham_single}
\hat{h}(\r)
= -\frac{\hbar}{2m}\nabla_{\r}^2 + V(\r) \,.
\end{equation}
The latter equation gives kinetic and potential energy constituents for an isolated particle in the trap $V(\r)$ and the second term of (\ref{eq:ham}) describes the inter-particle interaction of strength $\lambda_0$ and given by the potential $W(\r-\r') \equiv W(\R)$.
Importantly, here we work in dimensionless units where energy is measured in terms of $\hbar^{2}/(mL^{2})$.
We choose $\hbar=1$, set the particle mass to $m=1$ and the unit of length to $L=1\:\mathrm{\mu m}$.
This gives units of energy and time $E = \omega = 2\pi \times 116.26 \:\mathrm{Hz}$ and $T = 1.37 \:\mathrm{ms}$, respectively.

In this paper we focus on one-dimensional (1D) systems composed of $N=10$ bosons with contact interaction $W(R)=\delta(R)$.
The bosons are trapped by a harmonic potential with a Gaussian barrier:
\begin{equation}\label{eq:pot}
V(x)
= ax^2 + b\exp(-cx^2) \,.
\end{equation}

Let us discuss some properties of this trapping potential.
Obviously, for $b=0$ it reduces to a harmonic potential with frequency $\omega_{H} = \sqrt{2a}$.
Furthermore, for general $a,b,c \in \mathbb{R}^+$ and $k=bc/a>1$ the potential is a double-well with minima at $x_{1,2} = \pm \sqrt{c^{-1}\ln{k}}$.

\begin{figure}[!ht]
  \includegraphics[width=\linewidth]{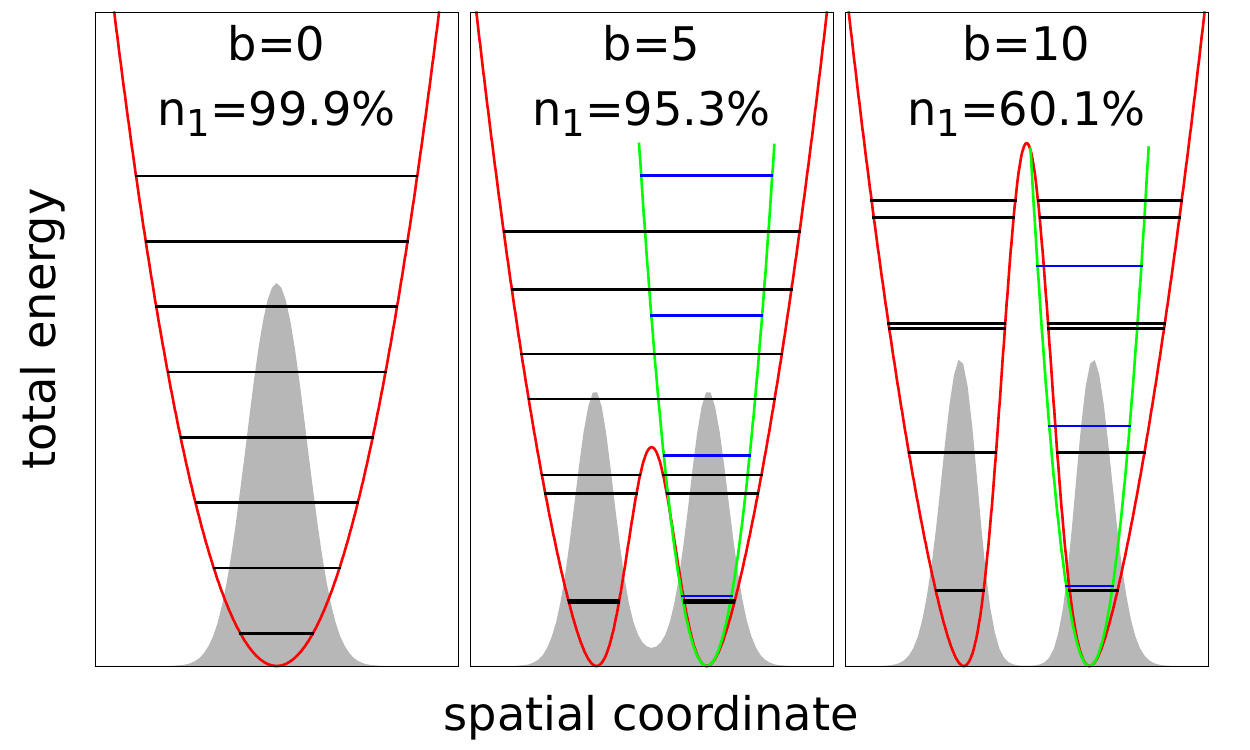}
  \caption{\label{fig:pots}
    \textbf{The three regimes studied:}
    Shapes of the trap potential (red lines), its eigenvalues (black lines) and wave-packet density (grey) are illustrated. The potentials (\ref{eq:pot}) are normalized as to vanish in the minima. For the double-wells the harmonic approximation at the potential minimum is shown (green potentials, blue eigenvalues). Occupation numbers of the first natural orbital $n_{1}$ and densities of the ground state are calculated for $N=10$ particles with $\Lambda\!=\!\lambda_0(N-1)\!=\!1$ at the MCTDHB($M=2$) level.
  }
\end{figure}

Typically, we set $a=1/2$ ($\omega_{H}=1$) and $c=1$ and distinguish between three major regimes according to the barrier height and the population of the most occupied (first) natural orbital:
i) a fully condensed BEC in a harmonic potential ($b=0$),
ii) a depleted BEC in a shallow double-well trap ($b=5$) and
iii) an almost fully fragmented BEC in a deep double-well potential ($b=10$).
The regimes are sketched in Fig.~\ref{fig:pots}.

Since the chosen potentials are symmetric to the origin, the Hamiltonian Eq.~(\ref{eq:ham}) is reflection invariant.
This allows a separation of the underlying Hilbert space into symmetric and an anti-symmetric manifolds.
We call states living in the symmetric and anti-symmetric subspaces \textit{gerade} ($g$) and \textit{ungerade} ($u$), respectively.

\section{\label{sec:results}Results}
\subsection{\label{sec:statics}Static picture}
In this section we analyze the excitation spectra of bosonic systems by virtue of the LR theory.
The procedure involves two steps ~\cite{grond13,LRuni,LRgeneralbench}.
First, we compute the ground state $E_{GS}$ for a given system using MCTDHB($M$) method.
Second, we compute the excitation spectrum by applying the LR-MCTDHB($M$) atop this static solution.
The number of orbitals $M$ determines the quality of the ground state.
Hence, the more orbitals $M$ used to describe the ground state, the greater the variety of excited states.
In particular, we distinguish between excitations that appear due to deformation of the involved orbitals $\phi_i$ (\textit{orbital-like}) and excitations that originate to the redistribution of the particles between given orbitals via configuration interaction (\textit{CI-like}) $C_{\vec{n}}$, see Eq.~(\ref{eq:Psi}) and Refs.~\cite{grond13,LRuni,LRgeneralbench} for more details.
For each excitation the response amplitude to a given perturbation ($\propto x$ or $\propto x^2$) is calculated.
Its magnitude determines the excited state's intensity in the respective excitation spectrum.
Physically it expresses the probability to excite the state by that specific perturbation.
For a detailed description of the LR-MCTDHB theory and its implementation see~\cite{grond13,LRuni,LRgeneralbench}.

We exploit the separability of the Hilbert space by calculating the response of the system to small perturbations of ungerade ($\propto x$) and gerade ($\propto x^2$) symmetry.
This allows us to uniquely identify each excitation from the ground state by its energy $\Delta E= E - E_{GS}$ and its symmetry (either $u$ or $g$).

\begin{figure}[!ht]
    \includegraphics[width=\linewidth]{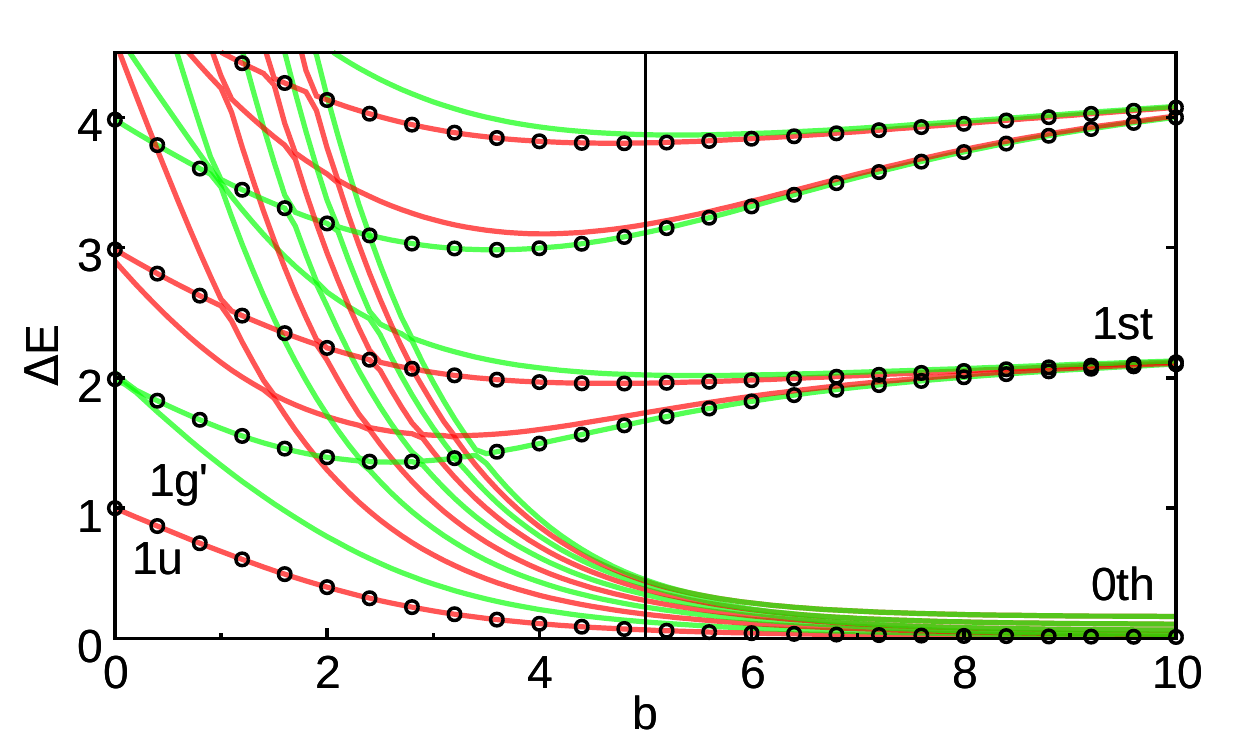}
    \caption{\label{fig:spec1}
        \textbf{Weak interaction:}
        Excitation spectrum as a function of barrier height $b$ calculated for $N=10$ bosons and $\Lambda=\lambda_0(N-1)\;=\;0.1$. $\Delta E$ is the excitation energy with respect to the ground state. Mean-field, LR-GP$\equiv$BdG, results are given by open black circles and many-body, LR-MCTDHB(2), predictions by colored lines. Green lines indicate \textit{gerade} and red lines \textit{ungerade} symmetry. States are labeled by their nodal structure and symmetry; primes are used to mark pure many-body excitations. See text for discussion.
    }
\end{figure}

In this work we consider the gradual transformation of a harmonic potential into a deep double-well potential by introducing a Gaussian barrier.
For weakly interacting bosons the excitation spectrum is shown in Fig.~\ref{fig:spec1}.
The standard LR-GP$\equiv$BdG$\equiv$LR-MCTDHB($M=1$) results are indicated by black circles and the many-body LR-MCTDHB($M\;=\;2$) results by colored lines, where green and red colors mark gerade and ungerade states, respectively.

From Fig.~\ref{fig:spec1} it is clear that there are states which are reproduced at both the mean-field 
LR-GP$\;\equiv\;$LR-MCTDHB($1$) and the many-body LR-MCTDHB($2$) levels.
However, one can also see states that are only present in the many-body treatment.
Following Refs.~\cite{grond13,LRuni,LRgeneralbench} we classify the obtained solutions according to this property further into \textit{mean-field-like} and \textit{many-body} excitations.

Let us discuss the differences between mean-field-like and pure many-body excitations.
For a non-interacting system the accessible states can be visualized in terms of particle-hole excitations involving eigenstates (orbitals) of the respective trap potential.
Within LR-GP (BdG) all bosons reside in the ground state and only $1h-1p$ excitations are allowed.
They describe the excitation of a single particle ($p$) to a higher-order one-particle state leaving a hole ($h$) in the ground state.
The order of a one-particle function is given by its nodal structure, where an even (odd) number of nodes corresponds to an excitation of gerade (ungerade) symmetry.
At the many-body level excitations may have a more complicated structure involving multiple particle excitations at the same time.

For the harmonic regime ($b=0$ in Fig.~\ref{fig:spec1}), excitations have been studied elsewhere~\cite{grond13,LRgeneralbench} by decoupling relative and center-of-mass (CoM) motion.
Essentially, there are mean-field-like and many-body states of equal symmetry that are energetically close.
As suggested there, we label these excitations according to their nodal structure and the symmetry they exhibit.
For example, $1u$ is the lowest-in-energy state with a one-node structure.
It describes the ungerade CoM excitation in a harmonic oscillator at $\Delta E=\omega_H=1$.
The higher-order $1h-1p$ states $2g,3u,\dots$ can correspond to the excitations of the CoM and relative motion.
Pure many-body states are indicated by a prime, e.g., $1g'$.
As shown in~\cite{grond13}, the lowest-in-energy many-body excitations correspond to higher-order CoM modes.
It is worthwhile to mention that in harmonically trap systems with a non-contact, e.g., parabolic interparticle interaction, the lowest-in-energy many-body states can correspond to the higher-order excitations of the relative motion ~\cite{LRgeneralbench}.

Next, we would like to discuss the almost fully fragmented regime in a deep double-well trap ($b=10$ in Fig.~\ref{fig:spec1}).
One clearly observes the formation of band-like structures for the asymptotic limit where the energy difference between each pair (green and red curves) vanishes.
In this limit of high barriers the entire excitation spectrum becomes two-fold degenerate and the ground state of the many-body system is two-fold fragmented.

In the deep double-well regime the LR-GP theory predicts only one excited state in the lowest band depicted in Fig.~\ref{fig:spec1} by open black circles.
This state corresponds to the $1h-1p$ excitation from the ground state GP orbital of the gerade symmetry to its complimentary ungerade one.
Similar 1h-1p excitations to higher quasi-degenerate excited orbitals form the mean-field description of the higher bands.
By contrasting these mean-field predictions with the LR-MCTDHB($M=2$) results plotted in Fig.~\ref{fig:spec1} by green and red lines we see that in the many-body description additional states contribute to the bands.
In the many-body picture the zeroth band is composed of ten states, equal to the total number of particles: $N=10$.
To understand this, let us once again return to the non-interacting case and recall that two lowest-in-energy eigenstates of the deep double-well trap are almost degenerate two-hump symmetric (\textit{gerade}) and an anti-symmetric (\textit{ungerade}) orbitals. 
In the non-interacting ground state all $N$ bosons occupy the gerade function.
The first single-particle excitation to the other orbital is described by $0u$, changing, thereby, the overall symmetry.
This $1h-1p$ excitation is the same as within the BdG theory.
Next excitation corresponds to the situation in which two particles are excited simultaneously from the ground state orbital to its anti-symmetric counterpart.
As a consequence the overall gerade symmetry of this many-body excited state marked as $0g'$ is re-established.
A three particle excitation process is described by $0u'$ and so on.
This branch of many-body excitations thereby corresponds to the successive population of the respective anti-symmetric orbital.
Formally, all these excitations can be described by different redistributions (configurations) of the particles among the two orbitals.
Therefore, these excitations are CI-like.

In a presence of interactions the many-body states of the lowest (0th) band become split.
However, this branch of excitations is hard to access, due to the involvement of multi-particle processes.
This physical conclusion is confirmed by the values of the corresponding linear response amplitudes, see Tables~\ref{tab:ES1}, \ref{tab:ES2} and \ref{tab:ES3} and their discussions.

In contrast to the lowest band, 
bands of higher energy at the LR-MCTDHB(2) level of description contain only four states.
Closer examination of the first band shows that two states are GP-like and the other two are pure many-body excitations.
Depending on the orbital in which the $1h-1p$ excitation is realized, the GP-like state is either of gerade or ungerade symmetry.
The corresponding many-body excited states have a $2h-1p-1p$ structure:
In addition to the GP-like excitation a second particle is excited to a higher-in-energy orbital.
At the LR-MCTDHB(2) level of description more complex excitation structures like $3h-2p-1p$, $4h-3p-1p$ are not reachable in higher bands.
This absence of the higher-order excitations reflects the fact that the MCTDHB(2) is an approximation to the exact solution and, therefore, higher-level theories are required for their proper descriptions, see e.g., discussion of the LR-MCTDHB(4) results.

\begin{figure}[!ht]
    \includegraphics[width=\linewidth]{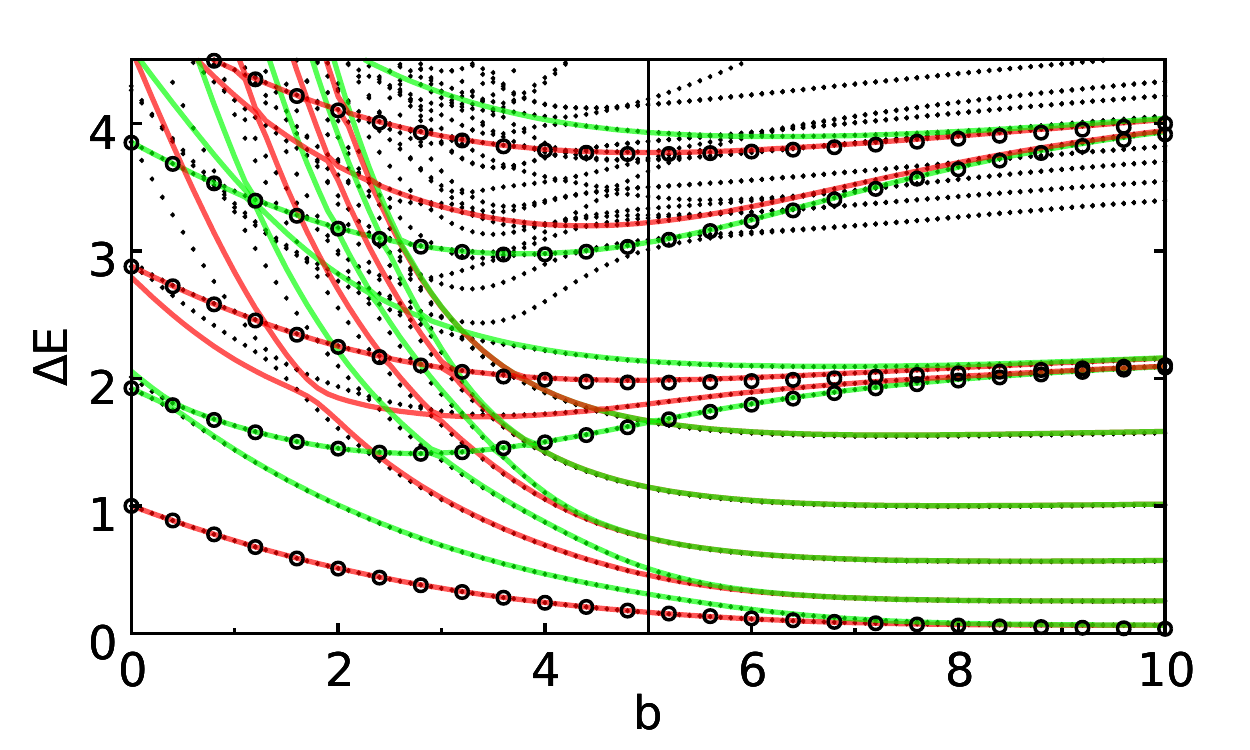}
    \caption{\label{fig:spec2}
        \textbf{Intermediate interaction:}
        Excitation spectrum as a function of barrier height $b$ for $N=10$ bosons and $\Lambda=\lambda(N-1)=1$. $\Delta E$ is the excitation energy with respect to the ground state. Open black circles depict the mean-field LR-GP$\equiv$BdG results. The two-orbital LR-MCTDHB($M=2$) many-body results are plotted by colored lines, four-orbital LR-MCTDHB($M=4$) many-body results -- by black dots. Green lines indicate \textit{gerade} and red lines \textit{ungerade} symmetry. The low-energy excitations are seen to converge. See text for the discussion.
    }
\end{figure}

For dynamic studies the clustering and degeneracy of states in bands complicates identification and attribution of the distinct excitations.
One way to overcome this issue is to investigate BECs in a shallow double-well ($b=5$ in Fig.~\ref{fig:spec1}), where the bands are not yet formed.
Another possibility to split the excited states is to increase the interparticle interaction strength.
In Fig.~\ref{fig:spec2} we show the same study as before but for stronger interparticle interaction strength $\Lambda = \lambda(N-1)=1$.
For convenience, we keep the introduced labeling.

Let us compare and contrast the excitation spectra at weak (Fig.~\ref{fig:spec1}) and intermediate (Fig.~\ref{fig:spec2}) interparticle interactions.
At the mean-field level (BdG - black circles) the overall spectra are quite similar.
The mean-field excitation energies are slightly shifted due to interparticle interaction.
All these mean-field excitations are reproduced at the many-body level (LR-MCTDHB($2$) - color lines) in both cases.
The main difference to the weak interaction, however, is a splitting of states in the zeroth band in the deep double-well regime.
At intermediate interactions the ultracold systems become less coherent i.e., more correlated and a single wave-function description is definitely insufficient.

By contrasting LR-MCTDHB($2$) and LR-MCTDHB($4$) (black dots) results depicted in Fig.~\ref{fig:spec2} we see that, as expected, at higher energies the four-orbital theory introduces additional many-body states and shifts positions of some excited states.
The observed shifts suggest that an even higher-level MCTDHB theory with more than $M=4$ orbitals is needed to converge the respective many-body states.
However, for the excitation energies smaller than $\Delta E \lesssim 3$ and not to small barriers ($b \gtrsim 4$) the deviations between two- and four-orbital LR-MCTDHB results are negligible.
We conclude that the usage of the two-orbitals LR-MCTDHB($2$) theory is sufficient to describe the low-lying excitations for all the here studied systems and in all the studied regimes.

We would like to mention that with stronger interaction strength \textit{avoided crossings} become visible.
They reflect the fact that states of same symmetry do not cross but rather exchange properties.
An example is the avoided crossing at $b\approx 2$ and $\Delta E \approx 2$ in Fig.~\ref{fig:spec2}.

To conclude the static part, the LR-MCTDHB($M$) theory has been used to  compute and study systematically excitation spectra of the interacting bosonic systems trapped in different external potential.
By introducing a Gaussian barrier $b$ the harmonic potential has been gradually transformed into a deep double-well potential.
The computed excitation energies $\Delta E$ and intensities $I$ for the three important regimes ($b=0,5,10$) are listed in Tables~\ref{tab:ES1}, \ref{tab:ES2} and \ref{tab:ES3}.
Next, we plan to contrast these static results with the dynamic studies in the following sections.

\begin{figure}[!htb]
  \includegraphics[width=\linewidth]{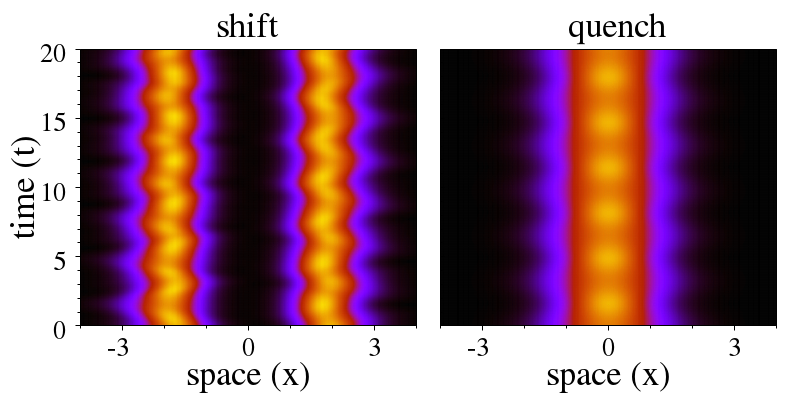}
  \caption{\label{fig:minkowski}
    \textbf{Propagation:}
    Right panel: Breathings in a harmonic potential induced by a quench of $a$ from $0.4$ to $0.5$. Left panel: Dipole-like oscillation in a deep double-well after shifting the potential by $x_{shift}=0.1$.  Calculations are done at MCTDHB($M=2$) level for $N=10$ particles and $\Lambda=\lambda(N-1)=1$.
  }
\end{figure}

\subsection{\label{sec:dynamics}Dynamics}
In this section we solve the many-body Schrödinger equation, Eq.~(\ref{eq:TDSE}), using MCTDHB($M$) wave-packet propagation.
We use the following protocol to access excited states:
Firstly, we obtain a given system in its ground state.
Secondly, we manipulate the Hamiltonian Eq.~(\ref{eq:ham}) and propagate the above obtained  state in an altered environment for two basic scenarios, namely \textit{shift} and \textit{quench}.
In order to activate ungerade excitations, which correspond to the \textit{dipole}-like movement of the BEC in a harmonic trap we propose to shift the origin $x \to x+x_{shift}$ of the external potential in Eq.~(\ref{eq:pot}).
The gerade (\textit{breathing}-like) excitations of the BEC can be activated by quenching the frequency of the outer harmonic potential  $a$-parameter in Eq.~(\ref{eq:pot}).
These sudden manipulations of the Hamiltonian induce a non-equilibrium evolution of the system which manifests itself as oscillation patterns of the one-particle density and in the temporal evolution of other observables.
We shall see that the proper choice of observables is very crucial as each one is sensitive to different aspects of the excitations and, therefore, provides different information about them. 

Let us start our discussion on suitable observables with the particle density which is computed as the diagonal part of the reduced one-body density Eq.~(\ref{eq:dns}):
\begin{equation}
\rho(x,t)
\equiv \rho(x,x;t)
= \sum^M_{i=1} n_i(t)\: \phi^{\ast NO}_i(x,t)\: \phi^{NO}_i(x,t) \,.
\end{equation}

In Fig.~\ref{fig:minkowski} the typical evolution of $\rho(x,t)$ in the shift and quench scenarios is shown.
Here, time is plotted against space ($x$ coordinate) in a Minkovskii-like manner and the particle density $\rho(x,t)$ is indicated by a color map:
Yellow and black colors mark regions of high and low density, respectively.
The left panel shows the \textit{dipole-like} movement of a BEC after shifting the origin of the double-well potential.
The right panel shows typical evolution of the system in the harmonic trap activated by the quench scenario.
The visible symmetric deformation of the density is known as a breathing mechanism~\cite{edwards96,mcdonald13}.

In addition to the particle density, we also analyze evolution of the expectation values of several one- and two-body operators.
The first one is the position operator defined for $N$-particle case as  $\hat x = \sum^N_{i}x_i$.
Its expectation value reads:
\begin{equation}\label{eq:x}
\langle x \rangle
= \sum^M_{i,j} \rho_{ij}(t)\: x_{ij} \,,
\end{equation}
where $\rho_{ij}$ are elements of the reduced density matrix, see Eq.~(\ref{eq:dns}) and $x_{ij} = \int  \phi^*_j(x,t)\: x\: \phi_i(x,t) \mathrm{d}x$.
The second one is the square of the position operator, which contains contributions from one- and two-body operators $\hat x^2 = (\sum^N_{i}x_i)^2 = \sum^N_{i} x^2_i + \sum^N_{i,j}x_ix_j$.
So, computation of its expectation value is a more involved task~\cite{klaiman15}:
\begin{equation}\label{eq:xx}
\langle x^2 \rangle
= \sum^M_{i,j}\rho_{ij}(t)\: x^{1b}_{ij}
+ \sum^M_{i,j,k,l}\rho_{ijkl}(t)\: x^{2b}_{ijkl} \,,
\end{equation}
where $x^{1b}_{ij}=\int  \phi^*_j(x,t)\: x^2\: \phi_i(x,t)\: \mathrm{d}x $ and $x^{2b}_{ijkl}=\int  \phi^*_j(x_1,t)\: \phi^*_l(x_2,t)\: x_1x_2\: \phi_i(x_1,t)\:\phi_k(x_2,t)\: \mathrm{d}x_1\mathrm{d}x_2$.
Finally, we can compute the corresponding variance
\begin{equation}\label{eq:var}
\Var(x)
= \langle x^2 \rangle - \langle x \rangle^2 \,.
\end{equation}
All these expectation values are integral quantities and, therefore, contain \textit{global} information about the system's evolution:
$\langle x \rangle$ gives the mean position of the cloud and its variance $\Var(x)$ - the deviation from the mean. 

In all the defined above expectation values time dependency is assumed implicitly.
Let us now derive formal explicit expressions.
The time evolution of any initial state $\left|\Psi(t=0)\right>$  can be re-expressed in terms of the exact eigenstates $\left|\Psi_n\right>$ of the Hamiltonian
\begin{equation}\label{eq:init}
\left|\Psi(t)\right>
= \sum_{n}a_{n} e^{-i E_n t} \left|\Psi_n(t)\right> \,,
\end{equation}
where $a_n= \langle  \Psi(t=0) \left|\Psi_n\right>$.
The evolution of the expectation value of any operator $\hat O$, thus, reads
\begin{equation}
\langle \Psi(t) | \hat O | \Psi(t) \rangle
= \sum_{n,m}a_{n}a_{m}e^{-i (E_n-E_m) t}
  \langle \Psi_m | \hat O | \Psi_n \rangle \,.
\end{equation}
It is important to stress that along with the differences in energy between the ground and excited states -- excitation energies, it also contains contributions from the differences in energy between other involved excited states.
From now on for the sake of simplicity we refer to the contributions from these higher-order processes which do not involve the ground state as de-excitations and mark them correspondingly, see for example $2g'\leftrightarrow 2g$ de-excitation in Table~\ref{tab:ES1}.
Their probabilities depend on the initial wave-function via expansion coefficients $a_{n}$ in Eq.~(\ref{eq:init}) and on the particular integrals $\langle \Psi_m| \hat O | \Psi_n \rangle$.
A naive expectation is that contributions to the evolution from the de-excitations should be negligible if the activated dynamics keeps the system in a linear response regime.
Formally, such a regime can be reached if only $a_{0}$ is dominant in the expansion of the initial wave-packet Eq.~(\ref{eq:init}).
In the present study we would like to verify it and to find out which expectation values -- of the one- or two-body operators are more sensitive to the dynamics and, therefore, contain more information about the excitations and de-excitations.

Complimentary to the integral observables, we would also like to monitor the density at some fixed-point position $x_0$
\begin{equation}\label{eq:rx}
  \rho(x=x_0,t) \,,
\end{equation}
which by definition is a \textit{local} observable.
We expect that the integral quantities can be used to characterize the BEC as an entity and the local particle density $\rho(x=x_0,t)$ to grasp possible local features.
From this perspective the integral observable can be regarded as a weighted average of the local quantities computed at all possible positions.

Let us mention some obvious implications of our dynamic protocols to the above defined quantities of the interest.
Obviously, $\langle x \rangle = 0$ holds for the quench scenario in a symmetric trap potential and, therefore, $\Var(x) = \langle x^2 \rangle$.
In other words, quench scenario can address only gerade excitations.
In the shift scenario, in general $\Var(x)\neq 0$ and the ungerade excitations manifest themselves in oscillations of $\langle x \rangle$.
Due to the above symmetry arguments, we plan to use the mean position $\langle x \rangle$ to study ungerade excitations in the shift scenario and the variance $\Var(x)$ to study gerade excitations in the quench scenario.

As mentioned above, in the dynamic scenarios excitations manifest themselves in temporal oscillations of the many-body wave-function and observables.
Hence, we use the MCTDHB wave-function computed at each propagation time-step to calculate the corresponding local $\rho(x=x_0)$ and integral $\langle x \rangle$,$\Var(x)$ observables.
Next, by applying a subsequent Discrete Fourier Transformation (FT) to each evolving quantity we extract the dominating oscillation frequencies, the corresponding excitation energies $\omega\equiv \Delta\E$ and their intensities, see appendix for more computational details.
One of the primary goals of the present study is to contrast the static LR predictions obtained in the previous section with these dynamic FT-extracted excitations and intensities.

A delicate issue is the length of propagation.
Naturally, a long propagation time results in a large number of oscillation circles that allows for an accurate extraction of frequencies.
The numerical task is to assure a small error in propagation especially for large quenches and displacements.
For our purpose, we shift the trap origin to $x_{shift}=0.1$ and quench the trap frequency $a$ from $0.4$ to $0.5$ in Eq.~(\ref{eq:pot}).
We assume that these values are in a linear regime such that excitations are predictable by the LR theory.
See appendix for more information on quantitative characterization of the linear and non-linear regimes in the studied dynamic scenarios.

In the following we study the shift and quench scenarios for the three regimes introduced in Sec.~\ref{sec:intro}.

\subsubsection{Harmonic Trap}
We start by analysing the dynamic scenarios in the harmonic potential.
In Table~\ref{tab:ES1} the static LR predictions and dynamic results are contrasted directly.
Excitation energies $\Delta E$ and corresponding intensities $I$ (in brackets) are listed for the different theories and dynamic scenarios.
We label the excitations as described in Sec.~\ref{sec:statics}.

\begin{table*}[ht!]
\begin{ruledtabular}
\begin{tabular}{c | cc  cccc}
            &BdG            &LR-MCTDHB(2)
            &"shift" (GP)   &"shift" (MB)
            &"quench" (GP)  &"quench" (MB)\\
label       &$\Delta E     \quad (I)$   &$\Delta E     \quad (I)$
            &$\Delta E     \quad (I)$   &$\Delta E     \quad (I)$
            &$\Delta E     \quad (I)$   &$\Delta E     \quad (I)$\\

\hline

$1u$        &1.000 (1.00)   &1.000 (0.41)
            &1.000 (1.00\fm[1]\fm[3])   &1.000 (1.00\fm[1]\fm[3])
            &-              &-           \\
$2g$        &1.922 (0.99)   &1.920 (0.84)
            &-              &-
            &1.922 (1.00\fm[2],0.64\fm[3])  &1.919 (0.40\fm[2],0.79\fm[3])\\
$2g'$       &-              &2.055 (0.13)
            &-              &-
            &-              &2.050 (0.53\fm[2])\\
$3u'$       &-              &2.792 (0.02)
            &-              &-
            &-              &-           \\
$3u$        &2.878          &2.880 (0.27)
            &-              &-
            &-              &-           \\
$4g$        &3.848 (0.01)   &3.851
            &-              &-
            &3.848 (0.36\fm[3])  &3.847 (0.21\fm[3])\\
\\
\hline

$2g' \leftrightarrow 2g$  &-              &0.135
            &-              &-
            &-              &0.130 (0.04\fm[2])\\
\end{tabular}
\end{ruledtabular}

\footnotetext[1]{observed in $\langle x \rangle$}
\footnotetext[2]{observed in $\Var(x)$}
\footnotetext[3]{observed in $\rho(x=1)$}
\caption{\label{tab:ES1}
    \textbf{Excited states in the harmonic potential:}
    Static and dynamic results for excitation energies $\Delta E \le 4.0$ and normalized intensities $I$ (in brackets) are presented at mean-field (GP) and many-body (MB) levels. The labeling is as described in Sec.~\ref{sec:statics}. For the statics the intensities $I$ are normalized for ungerade ($u$) and gerade ($g$) manifolds separately. In the dynamics normalization is done by summing over the intensities of all constituting peaks. States with intensity smaller than one percent are omitted. We attribute the lowest-in-energy dynamic process with $\Delta E = 0.130$ to a higher-order $2g' \leftrightarrow 2g$ excitation (de-excitation) because only these two lowest static LR states $2g'$,$2g$ have appropriate energies $\Delta E_{2g'}-\Delta E_{2g} = 0.135$ and symmetry. Footnotes in the dynamic picture indicate the quantity in which the excitation is observed. The dynamics are computed for $x_{shift}=0.1$ and a quench $a=0.4\to0.5$ using $N=10$ and $\Lambda=\lambda(N-1)=1$.
}
\end{table*}

Before going into a detailed analysis note that all the dynamic excitations and de-excitations extracted from the evolution of the respective observables are predicted by the static LR-MCTDHB theory.
We show that de-excitation energies can be obtained within the static LR-MCTDHB approach by taking differences between appropriate excited states.  
The frequencies obtained within the static and dynamic simulations match nicely implying that it is indeed possible to excite corresponding states dynamically by the suggested protocols.
For a quantitative comparison of the static and dynamic results the FT-computed intensities have been normalized over all constituting states of equal symmetry.
It is seen that dynamic results and LR predictions share a general tendency:
Excitations that have large response amplitudes in the LR predictions tend to be easily excited by the dynamic scenarios.

\begin{figure}[ht!]
    \includegraphics[width=\linewidth]{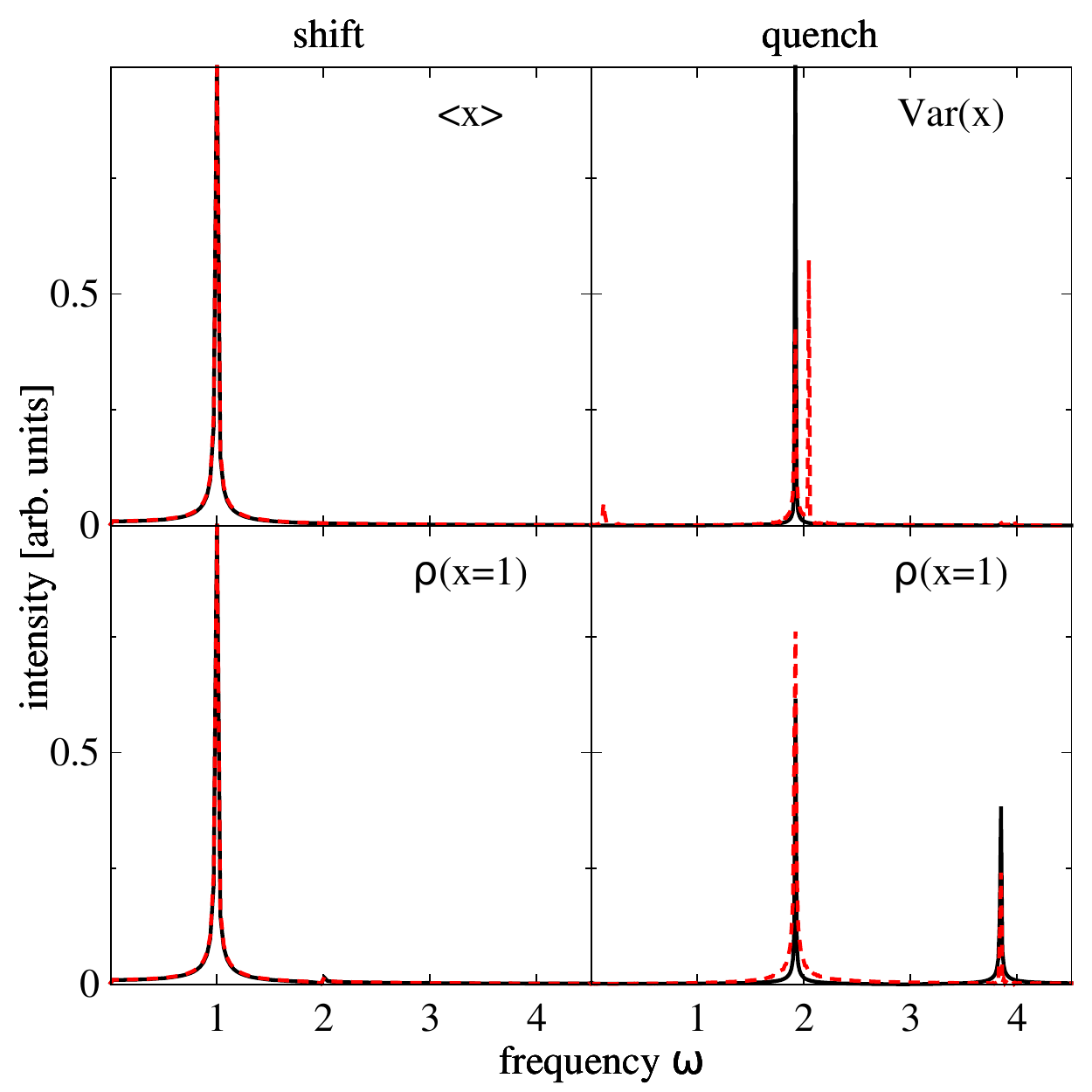}
    \caption{\label{fig:fft1}
        \textbf{Harmonic trap:}
        The left frame shows the frequency analysis of an oscillating BEC after a sudden shift $x_{shift}=0.1$ of the confining harmonic potential. The right frames shows the study of the quench scenario $a=0.4$ to $a=0.5$. The GP (black lines) and MCTDHB(2) (red dashed lines) results are shown. The intensities are normalized by the sum of all constituting peaks.
    }
\end{figure}

The two left panels of Fig.~\ref{fig:fft1} present a detailed analysis of the shift scenario in a harmonic trap.
They depict the Fourier spectra of the mean position $\langle x \rangle$ and a local density at fixed position $\rho(x=1)$.
Black lines plot mean-field MCTDHB($M=1$) results, red dashed lines show many-body MCTDHB($M=2$) results.
We directly see that $\langle x \rangle$ and $\rho(x=1)$ oscillate with a single frequency $\omega=1$ at both mean-field and many-body levels.
By comparison with the LR results (Table~\ref{tab:ES1}), we identify this frequency as the lowest-in-energy ungerade CoM excitation, labeled $1u$.
This result is independent of the magnitude of the applied shift and of the number $M$ of orbitals used in the static and dynamic MCTDHB($M$) simulations confirming thereby the separability of CoM and relative motion in harmonic traps.

The right panels of Fig.~\ref{fig:fft1} show the analysis of the quench scenario in the harmonic trap.
Now, the local density $\rho(x=1)$ and the variance  $\Var(x)$  which is a global quantity contain different amount of information.
Let us start by analyzing the variance.
The Fourier spectrum of $\Var(x)$ at the mean-field GP level (black line) reveals a single peak attributed to the lowest gerade excited state $2g$.
Within the many-body MCTDHB(2) treatment (red dashed lines) this picture changes drastically.
In addition to the $2g$ excitation a second peak corresponding to the pure many-body excitation $2g'$ is visible.
It is the most intense peak in the spectrum of the variance (see Table~\ref{tab:ES1}) stating that a many-body treatment is by all means inevitable for an accurate description of the dynamics.
The lowest-in-energy dynamic process seen as a small peak at low frequency with $\Delta E \approx 0.130$ can be attributed to a higher-order $2g' \leftrightarrow 2g$ excitation (de-excitation) because only these two lowest-in-energy states $2g'$,$2g$ in the static LR picture have appropriate symmetry and energetics $E_{2g' \leftrightarrow 2g} \equiv E_{2g'}-E_{2g}=\Delta E_{2g'}-\Delta E_{2g} = 0.135$.
The occurrence of this de-excitation might indicate that we are beyond the linear response regime.

The local density $\rho(x=1)$ shows a different picture than the integral observable $\Var(x)$.
On the one hand the many-body excitation $2g'$ is absent and on the other hand the higher-order mean-field excitation $4g$ becomes visible.
The former becomes clear when thinking of $2g'$ as an excitation involving several particles. 
Since $\rho(x=1)$ has a local character it simply cannot \textit{see} this global excitation.
However, this locality is the reason for the pronounced appearance of $4g$ corresponding to a higher mean-field excitation.
First, notice that the density $\rho(x=x_0)$ is high in regions close to the potential minimum $x_0=0$
and low in outer regions.
Second, recall that integral operators e.g., $\Var(x)$ yield the globally most intense excitations by integration (weighted sum) over the local densities $\rho(x=x_0)$ taken at all possible positions.
Since $4g$ is visible in $\rho(x=1)$ but not in $\Var(x)$, it gives negligible contribution to the global dynamics but important to the local dynamics.
From this observation we conclude that local quantities taken in the regimes of low densities can be more sensitive to higher excitations than the global ones.

The dynamic studies in the harmonic potential reveal that many-body excitations play a crucial role even in the case of a fully condensed system.
Many-body aspects of BECs, e.g., $2g'$ excitation can be observed in the proposed quench scenario by monitoring the evolution of the global operator $\Var(x)$.
We conclude that the variance being very informative measure of the many-body correlations ~\cite{klaiman15,klaiman16} also constitutes a sensitive probe for many-body excitations.
The fact that de-excitations are visible implies that we are out of the linear response regime, see appendix.
We also note that information about higher order mean-field excitations become visible in the local density $\rho(x=x_0)$ taken at regions of comparable low density.
In other words, local quantities can probe excitations that are negligible on a global scale.

\subsubsection{Shallow Double-Well}
\begin{figure}[ht!]
  \includegraphics[width=\linewidth]{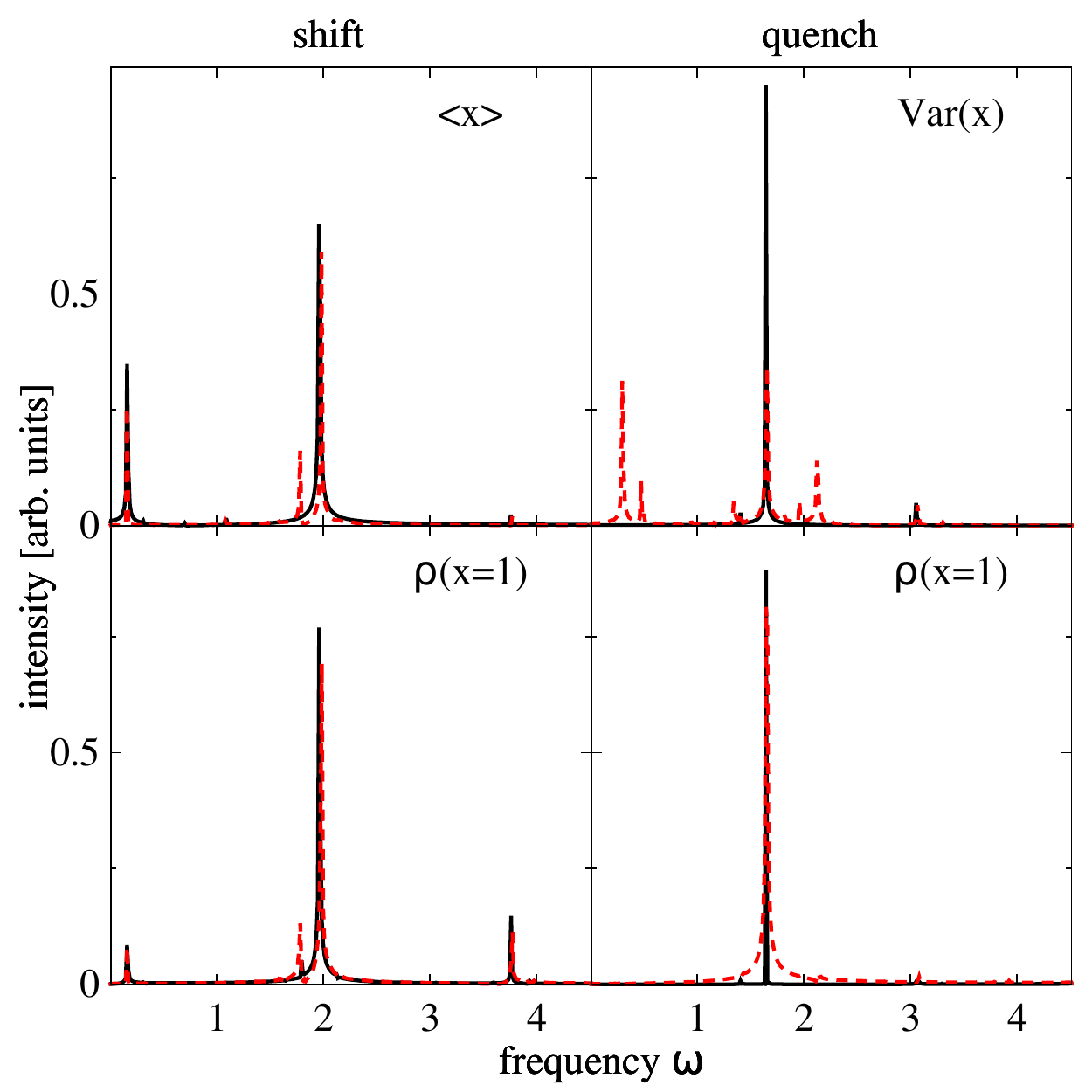}
  \caption{\label{fig:fft2}
    \textbf{Shallow double-well trap:}
    The description is as for Fig.~\ref{fig:fft1}.
  }
\end{figure}

As for the harmonic potential we analyze the integral $\langle x \rangle$, $\Var(x)$ and local $\rho(x=1)$ observables for the shift and quench scenarios now in the shallow double-well.
Figure~\ref{fig:fft2} depicts the discrete FT-spectra extracted from the evolution of these quantities.
The most striking observation is that a plethora of low-lying excitations become visible.
Here we would like to stress that the ground state of this system is slightly depleted  -- fragmentation ratio is $\approx5\%$.

\begin{table*}[ht!]
\begin{ruledtabular}
\begin{tabular}{c | cc  cccc}
            &BdG            &LR-MCTDHB(2)
            &"shift" (GP)   &"shift" (MB)
            &"quench" (GP)  &"quench" (MB)\\
label       &$\Delta E     \quad (I)$   &$\Delta E     \quad (I)$
            &$\Delta E     \quad (I)$   &$\Delta E     \quad (I)$
            &$\Delta E     \quad (I)$   &$\Delta E     \quad (I)$\\

\hline
$0u$        &0.166 (0.64)   &0.165 (0.52)
            &0.166 (0.34\fm[1],0.08\fm[3]) &0.164 (0.25\fm[1],0.08\fm[3])
            &-              &-\\
$0g'$       &-              &0.308 (0.02)
            &-              &-
            &-              &0.305 (0.29\fm[2])\\
$0u'$       &-              &0.456 (0.03)
            &-              &-
            &-              &-\\
$1g$        &1.648 (0.74)   &1.654 (0.65)
            &-              &-
            &1.648 (0.93\fm[2],0.89\fm[3]) &1.654 (0.32\fm[2],0.91\fm[3])\\
$1u'$       &-              &1.799 (0.15)
            &-              &1.782 (0.15\fm[1],0.12\fm[3])
            &-              &-\\
$1u$        &1.966 (0.31)   &1.986 (0.25)
            &1.966 (0.64\fm[1],0.74\fm[3]) &1.983 (0.58\fm[1],0.69\fm[3])
            &-              &-\\
$1g'$       &-              &2.134 (0.09)
            &-              &-
            &-              &2.129 (0.13\fm[2])\\
$2g$        &3.057 (0.21)   &3.066 (0.18)
            &-              &-
            &3.056 (0.04\fm[2],0.05\fm[3])              &3.063 (0.05\fm[2],0.03\fm[3])\\
$2u$        &3.758 (0.04)   &3.770 (0.02)
            &3.760 (0.02\fm[1],0.14\fm[3]) &3.771 (0.02\fm[1],0.11\fm[3])
            &-              &-\\

\\
\hline
$1g'\leftrightarrow 1g$  &-              &0.480
            &-              &-
            &-              &0.476 (0.08\fm[2])\\
$1g \leftrightarrow 0g'$  &-              &1.346
            &-              &-
            &-              &1.350 (0.04\fm[2])\\
$2g \leftrightarrow 1g$   &1.409          &1.412
            &-              &-
            &1.404 (0.02\fm[2],0.02\fm[3])              &1.408 (0.01\fm[2])\\
$1u \leftrightarrow 0u$   &1.800          &1.821
            &1.798 (0.03\fm[3])              &-
            &-              &-\\
\end{tabular}
\footnotetext[1]{observed in $\langle x \rangle$}
\footnotetext[2]{observed in $\Var(x)$}
\footnotetext[3]{observed in $\rho(x=1)$}
\end{ruledtabular}
  \caption{\label{tab:ES2}
  \textbf{Excited states in the shallow double-well:}
  The structure of the table is as in Table~\ref{tab:ES1}. In addition low-lying excitations with intensities smaller than one percent are neglected. These are, in particular, higher-order excitations of the zeroth band, compare to Fig.~\ref{fig:spec1}.
}
\end{table*}

As seen in Table~\ref{tab:ES2} every excitation energy obtained in the dynamic study can be attributed and identified with the corresponding static LR result.
Here we adopt the labeling from the deep double-well as described in Sec.~\ref{sec:statics}.
It is worthwhile to mention that the static LR amplitudes computed as responses of the system to small perturbations of ungerade ($\propto x$) and gerade ($\propto x^2$) symmetry in the double-well traps can give only a rough estimate for the excitability of the corresponding states by the dynamic protocols.
For a more quantitative characterizations a proper modification of the perturbation operators is required, which is out of the scope of the present study.

Before going into details let us discuss several important aspects of the shift scenario in the double-well traps.
First of all there is no separation of the center-of-mass and relative motions, all these states are coupled and the shift scenario can excite them.
The individual wells constituting the double-well are asymmetric, contrast in the Fig.~\ref{fig:pots} the red and green curves depicting the double-well traps and the harmonic approximations to each of these wells correspondingly.   
So, the shift of the origin of the double-well can lead to the excitations of all local modes of the individual wells.

The left panels of Fig.~\ref{fig:fft2} show the studies in the shift scenario.
As can be seen, the many-body spectra include the GP results and introduce the additional many-body excitation with pronounced intensity attributed in the Table~\ref{tab:ES2} as $1u'$.
Surprisingly, this state is seen in the evolution of both the global operator $\langle x \rangle$ and local quantity $\rho(x=1)$.
The fact that $1u'$ is also visible in local $\rho(x=1)$ indicates that relative and CoM motion are indeed coupled as opposed to the harmonic case.

The right panels of Fig.~\ref{fig:fft2} show the FT analysis of the $\Var(x)$ and $\rho(x=1)$ in the quench scenario.
Concerning the variance, the many-body treatment in addition to the mean-field GP peaks reveals two quite intense many-body excitations $0g'$ and $1g'$, see Table~\ref{tab:ES2}.
The most-intense mean-field $1g$ peak and these two many-body excitations have comparable intensities stressing the importance of a beyond mean-field treatment in shallow double-wells.
We also find that multiple de-excitations contribute to the variance spectrum (see Table~\ref{tab:ES2}).
This can be explained by the fact that the applied quench and shift manipulations pump more energy into the final system in the double-well studies comparing with that in the harmonic case,  see the appendix for more information.
More pumped energy means that a larger number of the excited states are populated increasing thereby the probability of the de-excitations seen, e.g., in the evolution of the integral two-body observables.

The situation is different for the local quantity.
In the quench scenario the local density $\rho(x=1)$ does not provide information about the many-body excitations.
As one can see in the right lower panel of Fig.~\ref{fig:fft2} the mean-field GP excitation $1g$ is the main and only excitation available.
Since we measure $\rho(x=1)$ close to the density minimum, higher order excitations do not contribute significantly.
However, taking the density at regions away from the minimum, the $2g$ excitation starts to contribute to the local dynamics.
Concluding, the local density can be used for accessing many-body excitations but a proper choice of the local position is very crucial.

\subsubsection{Deep Double-Well}

\begin{table*}[ht!]
\begin{ruledtabular}
\begin{tabular}{c | cc  cccc}
            &BdG            &LR-MCTDHB(2)
			&"shift" (GP)   &"shift" (MB)
			&"quench" (GP)  &"quench" (MB)\\
label       &$\Delta E     \quad (I)$   &$\Delta E     \quad (I)$
            &$\Delta E     \quad (I)$   &$\Delta E     \quad (I)$
            &$\Delta E     \quad (I)$   &$\Delta E     \quad (I)$\\

\hline

$0u$        &0.035 (0.46)   &0.063 (0.19)
            &0.034 (0.02\fm[1])   &-
            &-              &-\\
$0g'$       &-              &0.064
            &-              &-
            &-              &0.066 (0.36\fm[2])\\
$1g$        &2.085 (0.91)   &2.092 (0.81)
            &-              &-
            &2.085 (1.00\fm[2],0.93\fm[3])   &2.092 (0.33\fm[2],0.89\fm[3])\\
$1u$        &2.102 (0.46)   &2.095 (0.57)
            &2.102 (0.94\fm[1],0.66\fm[3])   &2.095 (0.93\fm[1],0.70\fm[3])
            &-              &-\\
$1u'$       &-              &2.158 (0.08)
            &-              &2.163 (0.02\fm[1],0.02\fm[3])
            &-              &-\\
$1g'$       &-              &2.161 (0.09)
            &-              &-
            &-              &2.161 (0.12\fm[2])\\
$2g$        &3.912 (0.04)   &3.934 (0.03)
            &3.914 (0.04\fm[3])                &-
            &3.914 (0.07\fm[3])                &3.934 (0.07\fm[2],0.10\fm[3])\\
$2u$        &3.997 (0.07)   &3.943 (0.07)
            &3.997 (0.04\fm[1],0.23\fm[3])   &3.943 (0.03\fm[1],0.15\fm[3])
            &-              &-\\
$2u'$       &-              &4.029 (0.05)
            &-              &4.032 (0.02\fm[1],0.07\fm[3])
            &-              &-\\
$2g'$       &-              &4.039
            &-              &-
            &-              &-\\

\\
\hline
$1g' \leftrightarrow 1g$  &-              &0.069
            &-              &-
            &-              &\\
$2g \leftrightarrow 1g$   &1.827          &1.841
            &-              &-
            &-              &1.839 (0.01\fm[2])\\
$2u \leftrightarrow 1u$   &1.894          &1.848
            &1.895 (0.01\fm[3])              &1.849 (0.02\fm[3])
            &-              &-\\
$1g \leftrightarrow 0g'$  &-              &2.028
            &-              &-
            &-              &2.032 (0.02\fm[2])\\
\end{tabular}
\end{ruledtabular}
\footnotetext[1]{observed in $\langle x \rangle$}
\footnotetext[2]{observed in $\Var(x)$}
\footnotetext[3]{observed in $\rho(x=1)$}

\caption{\label{tab:ES3}
  \textbf{Excited states in the deep double-well:}
  The structure of the table is as in Table~\ref{tab:ES1}.
}
\end{table*}

Finally let us study the dynamics in the deep double-well potential depicted in the left panel of Fig.~\ref{fig:pots}. 
For this choice of trapping potential the ground state is almost fully fragmented ($\approx40\%$) and, therefore, not described within the standard GP mean-field theory.
We recall that the BdG approach is capable of predicting the linear responses only form a fully condensed state, so its applicability in the present case of the deep double-well and fragmented ground state is questionable, nevertheless, we provide the LR-GP result for the sake of completeness. 
Indeed, the mean-field and many-body approaches already vary significantly in the description of the energy of the first excited state $0u$.
Table~\ref{tab:ES3} shows that the corresponding excitation energy $\Delta E$ within static LR-MCTDHB(2) is almost two times the energy of the lowest-in-energy excitation obtained by the mean-field BdG theory.

\begin{figure}[ht!]
  \includegraphics[width=\linewidth]{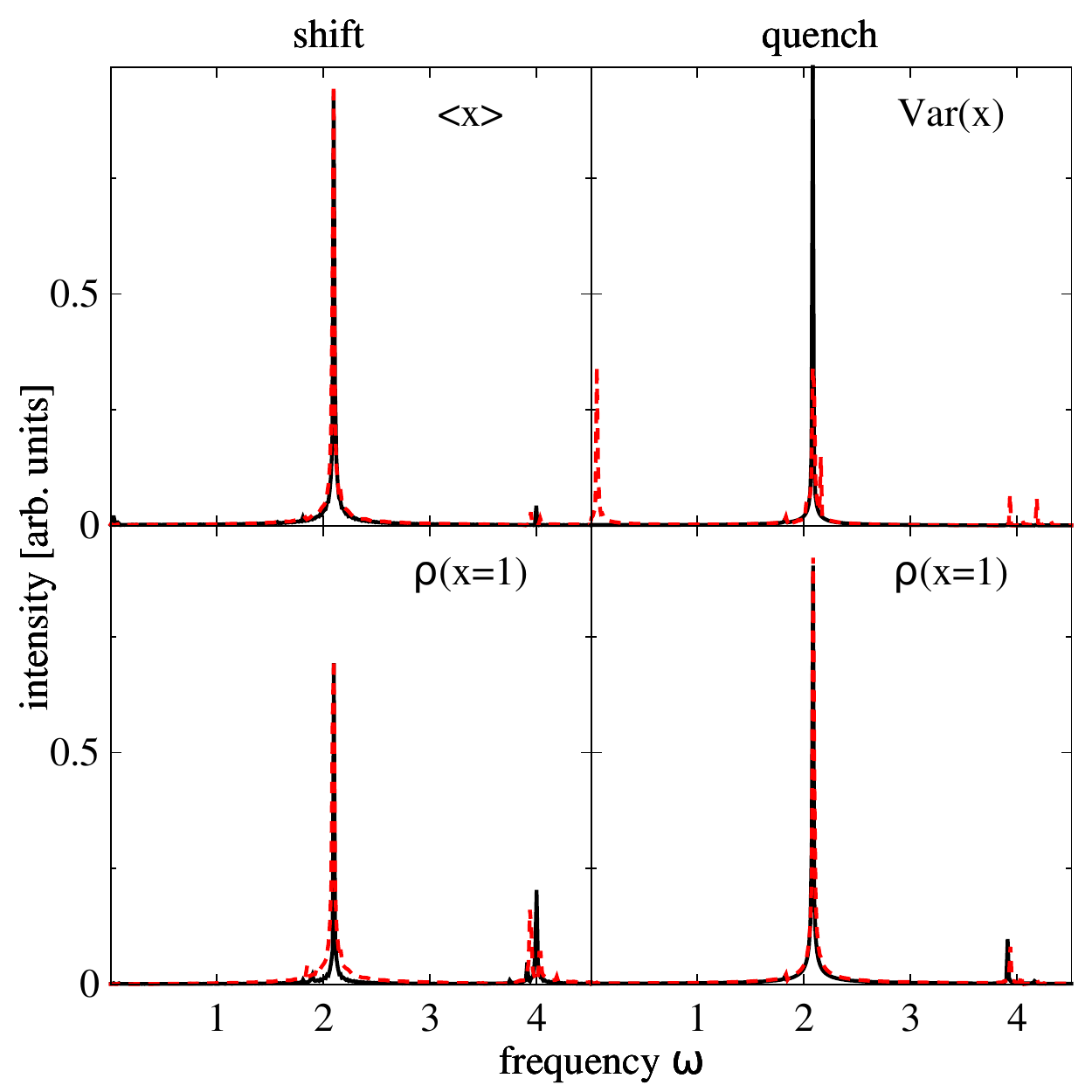}
  \caption{\label{fig:fft3}
    \textbf{Deep double-well trap:}
    The description is as for Fig.~\ref{fig:fft1}.
  }
\end{figure}

In Fig.~\ref{fig:fft3} we depict the discrete FT-results obtained from the evolutions of the local and integral observables after applying the quench and shift scenarios to the ground state of the deep double-well.
From Table~\ref{tab:ES3} it becomes clear that the overall situation is similar to the shallow double-well.
Most of the dynamic excitations can be identified with distinct LR states.
The main difference, however, is that excited states form bands, which result in more discrete spectra.

By analyzing the evolution of the $\langle x \rangle$ observable in the shift scenario we see that the $1u$ mean-field-like excitation contributes most to the dynamics as expected from the corresponding LR intensities.
This is true for both the mean-field and many-body descriptions.
The predicted many-body excitations $1u'$ and $2u'$ have quite small LR intensities and, therefore, do not contribute to the dynamics.
In the evolution of the local density $\rho(x=1)$ the higher order mean-field-like excitation $2u$ is more pronounced.
Still, in the shift scenario the integral and local observables give us the same picture of the excitations.
It is important to note that the lowest-in-energy excitation $0u$ is not visible although predicted by the LR intensities.
A possible explanation could be in a small energy difference between this and the ground state.
One has to monitor the evolution of the system for a longer propagation time in order to reach a higher resolution and detect contributions from this transition.

The right panels of Fig.~\ref{fig:fft3} show the excitation spectra obtained in the quench scenario of the deep double-well.
The overall situation is very similar to that observed in the shallow double-well.
Namely, the evolution of the variance $\Var(x)$ contains contributions from the excitations to the first ($1g$, $1g'$) and second ($2g$) bands.
However, the identification of the most intense peak at almost zero energy is not trivial, because energetically it can be attributed to the many-body excitation $0g'$ as well as to the inter-band de-excitation $1g' \leftrightarrow 1g$.
According to the LR results presented in the Table~\ref{tab:ES3}, the appearance of the $0g'$ excitation is rather unfavorable, because its LR intensity is negligible.
Also unlikely is that the de-excitation $1g' \leftrightarrow 1g$ constitutes the peak because its intensity in the comparable shallow double-well regime is small.
However, by monitoring the many-body evolution for a longer propagation time with the subsequent DFT analysis one should be able to split this lowest-in-energy peak into the contributing states and give further insight into the identification problem.
The local observable $\rho(x=1)$, similarly to the situation described in the shallow double-well, can access the mean-field-like $2g$ excitations to the second band, but not the low-lying many-body excitations.

The dynamic studies on the shallow and deep double-well potentials reveal one common feature of the local density $\rho(x=x_0)$.
This local quantity can be quite informative for detection of the many-body excitations in the shift scenario but not in the quench scenario.
Hence, the ungerade many-body excitations can in principle be measured by using local observables, however, to account for gerade many-body ones, it is desirable to use the many-body operators, such as $\Var(x)$.

\section{\label{sec:discussion}Summary}
In this paper we have investigated excitations and de-excitations of 1D BECs in the crossover from a fully condensed to a fully fragmented regimes.
The primary physical goal was to identify the nature and possible classes of the low-lying excitations and to compare and contrast predictions of the mean-field based (GP) and many-body (MCTDHB) theories.
On the methodological side we have employed two different techniques to access the excited states:
A static approach based on the linear response theories and a dynamic approach which utilizes the wave-packet propagation.

In the first section to compute the excitation spectra we have applied the static linear response mean-field BdG$\equiv$LR-GP and many-body LR-MCTDHB theories.
By comparing and contrasting the obtained results we have found that the many-body theory reproduces the excitations predicted at the mean-field level and, in addition, introduces new class of excitations which have pure many-body origin.
In particular, the low-lying many-body excitations have been observed in depleted and fragmented BECs confined in double-well potentials and, surprisingly, in condensed ultracold bosons trapped in harmonic traps.
In order to quantify the probability of these states to be excited by applied perturbations we have computed their linear response amplitudes. 
For interacting systems the many-body excitations have non-vanishing response amplitudes implying their relevance for quantum dynamics.
The comparable values of the response amplitudes obtained for the many-body and mean-field excited states imply that the many-body treatment is unavoidable for a proper description of these systems.
For further characterization of the excited states we have used their symmetry and nodal structure.

In the dynamic studies we have proposed two simple protocols that address excitations manifesting themselves in temporal oscillations of the many-body wave-function and observables.
We have investigated the dipole-like ungerade oscillations by shifting the origin of the confining potential and the breathing-like gerade excitations by quenching the frequency of the parabolic part of the trap.
The MCTDHB wave-function computed at each propagation time-step has been used to calculate the corresponding local $\rho(x=x_0)$ and integral $\langle x \rangle$,$\Var(x)$ observables.
By applying a subsequent Fourier transformation to each evolving quantity we have extracted the excitation energies and their intensities and compared them with the corresponding static linear-response predictions.
The main methodological conclusion is that all the dynamic excitations obtained within the studied protocols are predicted by and contained in the static LR-MCTDHB theory.
Moreover, the intensities of the FT-extracted excited states and the corresponding static LR predictions share a general tendency:
Excitations that have large response amplitudes in the LR predictions tend to be easily excited by the dynamic scenarios.
This holds true for mean-field and many-body excitations.

We have demonstrated that in the shift scenario the one-body observable $\langle x \rangle$ can be used to access the mean-field-like and many-body excitations.
The excitation probabilities of the mean-field-like states are, however, higher than that of the many-body states.
The two-body observable $\Var(x)$ used in the quench scenario is more informative quantity because along with the mean-field-like and many-body excitations it contains information about some de-excitations.
We have shown that these de-excitation energies can be obtained within the static LR-MCTDHB approach by taking differences between appropriate excited states.
The intensities of some pure many-body states are found to be large and comparable with mean-field-like states, implying that they can be directly detected.
We would like to mention that the local density taken at different regions of the trap can also be used to access excited states.
In particular, we have shown that the local densities taken at the regions of a comparatively low particle density, i.e., far from the potential minima can be used to access some higher order excitations which have not been seen in the global picture as given by $\langle x \rangle$ and $\Var(x)$.
However, these predictions are found to be very sensitive to the selected position and the dynamical scenario used.

Finally, in the present dynamic studies the quench and shift parameters of the Hamiltonian have been chosen such that the pumped energy was less than a few percents.
Nonetheless, we have observed small de-excitations contributing to the evolution of the local and global observables indicating the beyond the linear regime was reached.
As a consequence, in real experiments implementing simple dynamical scenarios it can be difficult to keep the system in the linear regime  and, therefore, a detailed knowledge of both excitations and de-excitations is crucial for a proper interpretation of the experimental data.

Alternatively, one can think about more sophisticated dynamical scenarios and control protocols where a major task would be to populate a desired excited state exclusively.
This kind of population control required for quantum computing, see e.g.~\cite{bason2012}, can, in principle, be reached by merging the optimal control algorithms with appropriate real-space many-body theories~\cite{OptControl1,OptControl2,OptControl3,OptControl4,CRAB-MCTDHB}.

\subsection*{Acknowledgments} 
We are indebted to O.E. Alon and S. Klaiman for fruitful scientific discussions.
Computation time on the bwGRiD  \cite{resbw-GRID}, and HybriLIT  \cite{resHybrilit} 
clusters as well as financial support by the DFG are greatly acknowledged.

\appendix*
\section{\label{app:LR}Linear Regime}
In this appendix we specify, quantify and determine the regimes of the dynamic changes of the Hamiltonian.
In the present studies all the dynamic manipulations on the system have been implemented by a sudden transition from an initial Hamiltonian $H_{i}$ to the final one $H_{f}$ involving, thereby,  \textit{finite} shifts and quenches.
This observation should be contrasted with the fact that the derivations of the LR methods are based on infinitesimal perturbations of the system.
Moreover, the term \textit{linear} in this respect specifies that only small first-order (linear) corrections induced by the applied perturbations to the many-body wave-function are taken into account and considered.

Hence we can formulate the problem as follow:
To what extend in a given dynamic scenario can we assure a linearity as required in the context of the LR approaches?

\begin{figure}[!ht]
    \includegraphics[width=\linewidth]{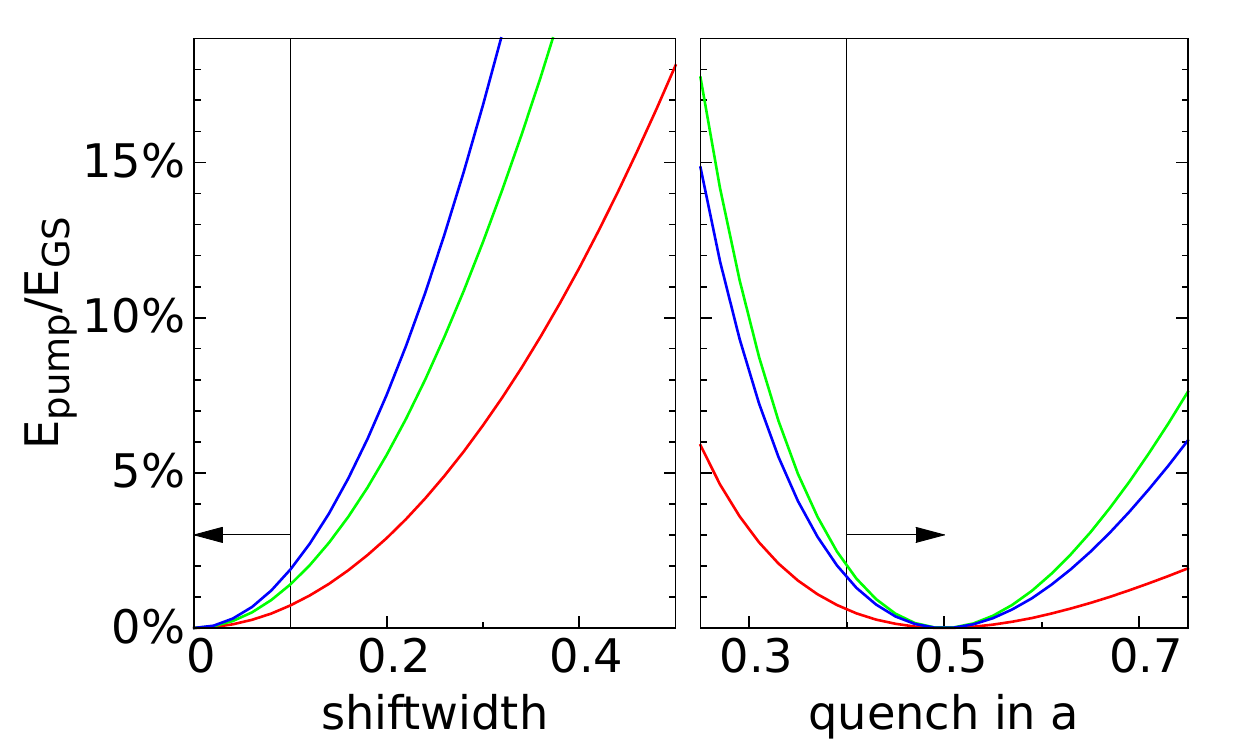}
    \caption{\label{fig:pump}
        \textbf{Pumped energy:}
        Energy pumped to the system by the proposed dynamic protocols which are based on the shift and quench of the trap potential, see Eq.~\ref{eq:pot} and its discussion. In the left panel, the pumped energy is plotted against the applied shift $x_{shift}$. The right panel shows the energy induced to the system by quenching the trap frequency $a$. Vertical lines and arrows denote the shift and quench parameters used in the paper. Red lines indicate results for the harmonic potential, green lines for the shallow double-well and blue lines for the deep double-well. Calculations are done at MCTDHB($M=4$) level for $N=10$ particles and $\Lambda=\lambda(N-1)=1$.
    }
\end{figure}

In order to answer this question let us study the energy which is pumped into a system by the applied dynamic protocols.
Let $E_{GS}$ be the ground state energy of the final system described by the $H_f$ and $E_{tot}$ be the total energy of the system after the applied dynamical scenario, i.e., after the sudden transition from $H_i \to H_f$.
The energy pumped into the system is then given by $E_{pump} = E_{tot} - E_{GS}$.
This amount of energy is redistributed between the excited states.
Figure~\ref{fig:pump} shows the pumped energy with respect to the ground state energy when the potential, given by Eq.~(\ref{eq:pot}), is shifted (left) or quenched (right).

Obviously, we have observed the non-linear behavior of the energy as a function of the presented control parameters in both cases.
So, for the studied systems the linear dependency of the pumped energy on the control parameter is unreachable in principle and, therefore, it cannot be considered as a criterion of the linear behavior.
Instead, we propose to rely on the magnitude of the pumped energy.
For our studies we choose shifts or quenches that pump a maximal amount of $3\%$ of the ground state energy into the systems.
This choice is practical, because it offers a compromise between two extremes.
On the one hand the magnitude of the pumped energy is large enough to be observed experimentally, while on the other hand non-linear effects like de-excitations are small but visible.

\section{\label{app:NumDet} Computational Details}
The derivation of the LR theory atop the MCTDHB wave-function is reported in Refs.~\cite{grond13,LRuni,LRgeneralbench}.
The final result for the resulting LR-MCTDHB theory boils down to the diagonalization of the non-hermitian LR matrix, i.e., takes on the form of the eigenvalue equation:

\begin{equation}\label{eq:LR_MCTDHB}
\bcalL 
\left(\!\!\begin{array}{c} 
\u^k \\ 
\v^k \\
\C_u^k \\ 
\C_v^k \\
\end{array}\!\!\right) =
\omega_k 
\left(\!\!\begin{array}{c} 
\u^k \\ 
\v^k \\
\C_u^k \\ 
\C_v^k \\
\end{array}\!\!\right).
\end{equation}

The linear-response matrix $\bcalL$ of the many-boson MCTDHB wave-function $\Psi$ is more involved than the commonly-employed Bogoliubov-de Gennes (BdG) linear-response matrix.
Physically, the response amplitudes of all modes, $\u^k$ and $\v^k$, and of all expansion coefficients, $\C_u^k$ and $\C_v^k$, combine to give the many-body excitation spectrum $\omega_k=E_k-E_{GS}$.
Here $E_{GS}$ and $E_k$ are the energies of the MCTDHB ground state and excited states respectively.
We have successfully managed to explicitly construct $\bcalL$ and obtained the many-body excitation spectrum for bosons interacting by contact ~\cite{grond13} and general ~\cite{LRgeneralbench} interparticle interaction potentials.

To quantify the intensity of the response we compute the response weights:
\begin{equation}\label{eq:weight}
\begin{array}{r@{}l}
I_k \equiv \gamma_k
&{}=
\langle\u^k| f^+(\brho^0)^{1/2} |\bphi^0\rangle
+ \langle\v^k| f^-(\brho^0)^{1/2} |\bphi^{0,*}\rangle \\
&{}
+ \left(\int \mathrm{d}\r\phi_i^{0,*}f^+\phi_j^0\right)
\langle\C_u^k|\hat a_i^{\dagger} a_j|\C^0\rangle
+ \left(\int \mathrm{d}\r\phi_i^{0,*}f^{-,*}\phi_j^0\right)
\langle\C_v^k|\left(\hat a_j^{\dagger} a_i\right)^*|\C^{0,*}\rangle \,.
\end{array}
\end{equation}
Here all quantities with 0-superscript relate to the static MCTDHB solution and  $f^{\pm}$ define the driving amplitudes of the applied perturbations, see Ref.~\cite{grond13} for details.

To identify the symmetry and nodal structure of the excited state we have computed and analyzed the real oscillatory parts of the corresponding response densities $\Delta\rho=\mathrm{Re}[ \Delta \rho^k_o(\r) + \Delta\rho^k_c(\r) ]$.
The contributing orbital and CI-terms are given as following 
\begin{equation}\label{eq:density_osc_orb}
\Delta\rho^{k}_o(\r)
= \sum_{i,j=1}^M \rho_{ij}^0\phi_i^0(\r)
  \left\{ \tilde u_j^k(\r) + \tilde v_j^{k}(\r) \right\} \,,
\end{equation}
where $\tilde u^k_j = (\brho^0)^{-1/2}_{ji} u^k_i$, and $\tilde v^k_j = (\brho^0)^{-1/2}_{ji} v^k_i$.
Further, we defined $\hat \rho^0=\sum_{ij}\hat a_i^{\dagger}\hat a_j\phi_i^{0,*}(\r)\phi_j^{0}(\r)$.
\begin{equation}\label{eq:density_osc_coef}
\Delta\rho^{k}_c(\r)
= \langle \C^0|\hat \rho^0|\C_u^k\rangle
+ \langle \C_v^{k,*}|\hat \rho^0|\C^0\rangle \,.
\end{equation}

In our time dependent studies for the temporal evolution of the quantity $f(t)$ given in time-slices $f(0),f(1),f(2),..,f(N-1)$ we compute its Discrete Fourier Transform (DFT) which is the equivalent of the continuous Fourier Transform for a time-evolving signal given only at $N$ instants (time-points) separated by equidistant sample times $\Delta t=\tau$ (i.e. a finite sequence of data):
\begin{equation}
F(i\omega)=\int^{(N-1)\tau}_0  f(t) e^{-i\omega t} dt \to \sum^{N-1}_{k=0} f(k)e^{-i k \omega \tau},
\end{equation}
or, more precisely with $\omega=0, \frac{2 \pi}{N \tau} \times 1, \frac{2 \pi}{N \tau}\times 2, \cdots, \frac{2 \pi}{N \tau}\times (N-1)$
\begin{equation}
\label{eq:evol_DFT}
F(n)=\sum^{N-1}_{k=0} f(k)e^{-i k \frac{2 \pi n }{N} }.
\end{equation}
We have normalized the obtained $F(n)$ amplitudes such that their sum would be equal to the unity.

In all our numerical simulations with the MCTDHB-Laboratory package~\cite{MCTDHB-Lab} we have use the Sin-DVR, Exp-DVR or FFT-grid in a box  $(-10:10)$ with $N_g=512$ grid points.
All the time propagations have been done till $T=500$.

\FloatBarrier

\end{document}